\documentclass{appolb}
\usepackage{graphicx}
\usepackage[usenames]{color} 
\newcommand{\beq}{\begin{equation}}
\newcommand{\eeq}{\end{equation}} 
\newcommand{\beqa}{\begin{eqnarray}}
\newcommand{\eeqa}{\end{eqnarray}} 
\newcommand{\ba}{\begin{array}}
\newcommand{\ea}{\end{array}}

\begin{document}

\title{Dimensional reduction and localization \\ 
of a Bose-Einstein condensate \\
in a quasi-1D bichromatic optical 
lattice\thanks{Presented at the 7th Workshop on Quantum Chaos 
and Localisation Phenomena, May 29-31, 2015 - Warsaw, Poland}}

\author{L. Salasnich$^{1,2}$ and S.K. Adhikari$^3$
\address{$^1$Dipartimento di Fisica ``Galileo Galilei'' and CNISM, \\
Universit\`a di Padova, Via Marzolo 8, 35131 Padova, Italy \\
$^2$Istituto Nazionale di Ottica (INO) del Consiglio Nazionale 
delle Ricerche (CNR), \\ 
Sezione di Sesto Fiorentino, Via Nello Carrara, 
1 - 50019 Sesto Fiorentino, Italy\\
$^3$Instituto de Fisica Teorica, 
Universidade Estadual Paulista \\ 
UNESP 01.140-070 Sao Paulo, Sao Paulo, Brazil}}

\maketitle

\begin{abstract} 
We analyze the localization of a Bose-Einstein condensate (BEC) 
in a one-dimensional bichromatic quasi-periodic optical-lattice potential 
by numerically solving the 1D Gross-Pitaevskii equation (1D GPE). 
We first derive the 1D GPE from the 
dimensional reduction of the 3D quantum field theory 
of interacting bosons obtaining two coupled differential 
equations (for axial wavefuction and space-time dependent 
transverse width) which reduce to the 1D GPE under 
strict conditions. Then, by using the 1D GPE 
we report the suppression of localization 
in the interacting BEC when the repulsive scattering 
length between bosonic atoms is sufficiently large. 
\end{abstract}

  
\section{Introduction}

Many years ago Anderson \cite{anderson} predicted 
the localization of the electronic wave 
function in a disordered potential. In the last twenty years 
the phenomenon of localization due to 
disorder was experimentally observed in electromagnetic 
waves \cite{light,micro}, in sound waves \cite{sound}, 
and also in quantum matter waves \cite{billy,roati,chabe,edwards}. 

In the case of quantum matter waves, 
Roati {\it et  al.} \cite{roati} observed localization of a 
non-interacting Bose-Einstein condensate (BEC) of $^{39}$K atoms 
in a 1D potential created by 
two optical-lattice potentials with different amplitudes 
and wavelengths. The non-interacting BEC of 
$^{39}$K atoms was created \cite{roati} by tuning the inter-atomic 
scattering length to zero near a Feshbach resonance \cite{fesh}. 
The 1D quasi-periodic potentials have a spatial ordering that is 
intermediate between periodicity and disorder \cite{harper,aubry,thouless}. 
In particular, the 1D discrete Aubry-Andre model of quasi-periodic 
confinement \cite{aubry,thouless} displays a transition from extended 
to localized states which resembles the Anderson localization 
of random systems \cite{random,optical}. Modugno \cite{modugno} 
has recently shown that the linear 1D Schr\"odinger equation 
with a bichromatic periodic potential 
can be mapped in the Aubry-Andre model 
and he studied the transition to localization 
as a function of the parameters of the periodic potential. 
 
To investigate the interplay between the bichromatic potential 
and the inter-atomic interaction in the localization of a BEC, 
we use the 1D Gross-Pitaevskii equation (1D GPE) \cite{book}. 
We first show that the 1D GPE can be derived 
from the dimensional reduction of the 3D quantum field theory 
of interacting bosons \cite{sala-barb1}, obtaining a 
nonpolynomial Schr\"odinger equation \cite{sala-npse,ioboriseflavio} 
which reduces to the 1D GPE only under 
strict conditions. Then we numerically solve the 1D GPE by 
using a Crank-Nicolson 
predictor-corrector method. We find that the cubic 
nonlinearity of the 1D GPE, which accurately 
models the binary inter-atomic interaction of atoms,  
has a strong effect on localization: a reasonably weak repulsive nonlinear 
term is capable of destroying the localization \cite{sala-loc}. 
Our results on the  effect of nonlinearity in the 
localization are in qualitative agreement with similar predictions 
based both 1D continuous \cite{addit} and discrete nonlinear 
Schr\"odinger equation with random on-site energies \cite{dnlse}. 

We stress that, in addition to the results discussed here on a pure BEC 
in a bichromatic lattice (see also \cite{sala-loc}), 
in the last few years there have been studies of localization 
of a BEC vortex \cite{sa1}, 
of a dipolar BEC \cite{sa2}, of a spin-orbit coupled BEC \cite{sa3}, 
of a Bose-Fermi mixture \cite{sa4}, of a BEC on a random 
potential \cite{sa5}, among other possibilities of 
localization of matter wave. 

\section{1D Gross-Pitaevskii equation with a quasi-periodic 
bichromatic potential}

In the experiment of Roati {\it et al.} \cite{roati}
the 1D quasi-periodic bichromatic optical-lattice  potential was 
produced by superposing two optical-lattice potentials generated by two 
standing-wave polarized laser beams of slightly different wavelengths 
and amplitudes. With  a single periodic potential 
the linear Schr\"odinger equation permits only 
delocalized states in the form of Bloch waves. Localization is possible 
in the linear Schr\"odinger equation due to the ``disorder'' introduced 
through a second periodic component. 

We model the dynamics of a trapped BEC of $N$ atoms 
in a transverse harmonic potential of frequency $\omega_{\bot}$
plus the axial quasi-periodic optical-lattice potential 
by using the following adimensional 1D Gross-Pitaevskii 
equation (1D GPE) \cite{book,sala-loc}
\begin{eqnarray}
\label{gp}
i\frac{\partial }{\partial t} \phi(z,t) = 
\biggr[-{1\over 2}{\partial_z^2}+ V(z)
+g|\phi(z,t)|^2\biggr]\phi(z,t)  \; ,  
\end{eqnarray}
where 
\beq
V(z)= \frac{4\pi^2 s_1}{\lambda_1^2}
\cos^2\biggr(\frac{2\pi}{\lambda_1}z\biggr) + 
\frac{4\pi^2 s_2}{\lambda_2^2}
\cos^2\biggr(\frac{2\pi}{\lambda_2}z\biggr) \; 
\label{pot1} 
\eeq
is the quasi-periodic bichromatic axial potential, with $\phi(z,t)$ 
the axial wave function of the Bose condensate normalized to one, i.e. 
\beq 
\int_{-\infty}^{\infty} dz |\phi(z,t)|^2 =1 \; . 
\eeq
Here $g=2Na_s/a_{\bot}$ is the dimensionless interaction strength 
with $a_s$ the inter-atomic scattering length and 
$a_{\bot}=\sqrt{\hbar/(m\omega_{\bot})}$ the characteristic 
harmonic length of the transverse harmonic confinement \cite{sala-loc}. 
Moreover, $2s_i, i=1,2,$ are the amplitudes  of the 
optical-lattice potentials in units of 
respective recoil energies $E_i=2\pi^2 \hbar^2/(m \hat \lambda_i^2)$, and 
$k_i=2\pi/\lambda_i$, $i=1,2$ are the respective wavenumbers, $\hbar$ 
is the reduced Planck constant, and $m$ the mass of 
an atom \cite{roati,sala-loc}. 
The optical potential of wavelength $\lambda_1$ is used to create 
a primary lattice that is weakly perturbed by a secondary lattice 
of wavelength $\lambda_2$ \cite{roati,modugno}. 
Moreover, to obtain ``quasi-disorder'' 
the ratio $\lambda_2/\lambda_1$ should not be 
commensurable \cite{modugno}. 
In practice we use $\lambda_2/\lambda_1=0.86$ that is close to 
the experimental value $\lambda_2/\lambda_1=0.835$ \cite{roati,sala-loc}. 

\section{Derivation of the 1D Gross-Pitaevskii equation}

Before solving the 1D GPE, let us analyze its derivation from the 
quantum theory of many-body systems \cite{book,sala-book}
The quantum many-body Hamiltonian of interacting identical 
bosons is given by 
\beqa
{\hat H} &=& \int d^3{\bf r} \
{\hat \psi}^+({\bf r})
\left[ -{\frac{1}{2}}\nabla^{2}
+ U({\bf r})  \right] {\hat \psi}({\bf r}) 
\nonumber 
\\
&+& 
\int d^3{\bf r} \ d^3{\bf r}' \
{\hat \psi}^+({\bf r})
{\hat \psi}^+({\bf r}') 
W({\bf r},{\bf r}') 
{\hat \psi}({\bf r}') {\hat \psi}({\bf r})
\label{ham}
\eeqa
where ${\hat \psi}(\mathbf{r},t)$ is the bosonic field operator. 
In our case the external trapping potential reads 
\beq 
U({\bf r}) = {1\over 2} (x^2 + y^2) + V(z) \; , 
\label{pio1}
\eeq
corresponding to a harmonic transverse confinement 
of frequency $\omega_{\bot}$ with characteristic 
length $a_{\bot}=\sqrt{\hbar/(m\omega_{\bot})}$ 
and the axial optical lattice $V(z)$ of Eq. (\ref{pot1}). 

In addition, due to the fact that the system is made of 
dilute and ultracold atoms, we consider a contact 
interaction between bosons, i.e. 
\beq
W({\bf r}-{\bf r}') = \gamma \, \delta^{(3)}({\bf r}-{\bf r}') 
\label{pio2}
\eeq
with $\delta^{(3)}({\bf r})$ the Dirac delta function and
\beq
\gamma = 2 {a_s\over a_{\bot}}
\eeq
the adimensional strength of the boson-boson interaction,
proportional to the s-wave scattering length $a_s$
of the inter-atomic potential $W({\bf r},{\bf r}')$.

Taking into account Eqs. (\ref{pio1}) and (\ref{pio2}), 
the Heisenberg equation of motion of the field operator
\beq
i {\partial \over \partial t} {\hat \psi}
= [ {\hat \psi} , {\hat H} ] \;
\eeq
gives
\beq
i{\frac{\partial}{\partial t}}{\hat \psi}(\mathbf{r},t) =
\left[ -{\frac{1}{2}}\nabla^{2}
+ {1\over 2} \left( x^2 + y^2 \right)
+ V(z) + 2\pi \gamma {\hat \psi}^+(\mathbf{r},t)
{\hat \psi}(\mathbf{r},t) \right] {\hat \psi}(\mathbf{r},t) \; . 
\label{quantum-3dgpe}
\eeq

In the superfluid regime, where the many-body quantum 
state $|QS\rangle$ of the system can be approximated by 
a Glauber coherent state $|CS\rangle$ of ${\hat \phi}(z)$ 
\cite{sala-barb1,sala-book}, i.e. such that 
\beq 
{\hat \psi}({\bf r},t) |CS\rangle = \psi({\bf r},t) |CS\rangle \; , 
\eeq
the Heisenberg equation of motion (\ref{quantum-3dgpe}) becomes 
the familiar 3D Gross-Pitaevskii equation (3D GPE) \cite{book} 
\beq
i{\frac{\partial}{\partial t}}{\psi}(\mathbf{r},t) =
\left[ -{\frac{1}{2}}\nabla^{2}
+ {1\over 2} \left( x^2 + y^2 \right)
+ V(z) + 2\pi \gamma |{\psi}(\mathbf{r},t)|^2 \right] 
{\psi}(\mathbf{r},t) \; , 
\label{3dgpe}
\eeq
where $\psi({\bf r},t)$ is a complex wavefunction normalized to the total 
number $N$ of bosons, i.e. 
\beq 
\int d^3{\bf r} |\psi({\bf r},t)|^2 = N \; . 
\eeq 
The time-dependent 3D GPE (\ref{3dgpe}) is the Euler-Lagrange equation 
of the action functional 
\beq 
S = \int dt \, d^3{\bf r} \ {\cal L} 
\eeq 
with Lagrangian density 
\beq 
{\cal L} = \psi^* \left[ i{\frac{\partial}{\partial t}} + 
{\frac{1}{2}}\nabla^{2} \right] \psi 
- {1\over 2} \left( x^2 + y^2 \right) |\psi|^2 
- V(z) |\psi|^2 - \pi \gamma |\psi|^4 \; . 
\eeq

To perform the dimensional reduction we suppose that 
\beq
{\psi}({\bf r},t)  = 
{N^{1/2}\over \pi^{1/2} \sigma(z,t)}
\exp{\left[ - \left( {x^2+y^2\over 2\sigma(z,t)^2} 
\right) \right] }\, {\phi}(z,t) \; , 
\label{qassume}
\eeq
where $\sigma(z,t)$ and ${\phi}(z,t)$ account respectively 
for the transverse width and for the axial bosonic 
wavefunction. We apply this Gaussian ansatz to the action functional  
of the 3D GPE \cite{sala-gaussian}. Integrating over $x$ and $y$ 
and neglecting the derivatives of $\sigma(z,t)$, we obtain the 
effective 1D action \cite{sala-npse}
\beq
{S}_{e} = \int dt \, dz \ {\cal L}_e 
\eeq
with the effective 1D Lagrangian density \cite{ioboriseflavio}
\beq 
{\cal L}_e = 
{\phi}^* \Big[ i {\partial \over \partial t} 
+{\frac{1}{2}}\partial_z^{2} \Big]  {\phi} 
 - V(z) |\phi|^2 - {1\over 2} 
\left( {1\over \sigma^2} + \sigma^2 \right) |\phi|^2 
- {\left({\partial_z\sigma}\right)^2 \over 2\sigma^2} |\phi|^2  
- {g\over 2\sigma^2} |\phi|^4 \; ,  
\label{qe1}
\eeq
and 
\beq 
g=N\gamma = {2N a_s\over a_{\bot}} \; . 
\eeq
Calculating the Euler-Lagrange equations of both $\phi(z,t)$ 
and $\sigma(z,t)$ one gets \cite{sala-npse,ioboriseflavio}
\beq 
i {\partial \over \partial t} \phi = 
\Big[ -{\frac{1}{2}}\partial_z^{2} + V(z) + {1\over 2} 
\left( {1\over \sigma^2} + \sigma^2 \right) 
+ {g\over \sigma^2} \ |\phi|^2 \Big] \phi \; , 
\label{1dnpse-1}
\eeq
and 
\beq 
\sigma^4 = 1 + g |\phi|^2 + (\partial_z \sigma)^2 
+ {\sigma^3\over |\phi|^2} \partial_z 
\left( {\partial_z\sigma\over \sigma^2} |\phi|^2\right) \; . 
\eeq
Neglecting the spatial derivatives of $\sigma(z,t)$ (adiabatic 
approximation) the last equation becomes \cite{sala-npse}
\beq 
\sigma = \left( 1 + g |\phi|^2 \right)^{1/4} \; .  
\label{1dnpse-2}
\eeq
Eqs. (\ref{1dnpse-1}) and (\ref{1dnpse-2}) give the 1D nonpolynomial 
Schr\"odinger equation (NPSE) 
\beq 
i {\partial \over \partial t} \phi = 
\Big[ -{\frac{1}{2}}\partial_z^{2} + V(z) + {1\over 2} 
\left( {1\over\sqrt{1 + g |\phi|^2}} + \sqrt{1 + g |\phi|^2} \right) 
+ {g|\phi|^2\over \sqrt{1 + g |\phi|^2}} \Big] \phi \;  
\label{npse}
\eeq
we introduced some years ago \cite{sala-npse}.
Finally, only under the condition 
\beq 
g |\phi(z,t)|^2 \ll 1 
\eeq
one finds $\sigma =1$ (i.e. $\sigma=a_{\bot}$ is dimensional units) 
and Eq. (\ref{1dnpse-1}) becomes 
the 1D GPE, Eq. (\ref{gp}). Notice that NPSE, Eq. (\ref{npse}), 
has been used by many authors to study quasi-1D BECs with 
a transverse width $\sigma$ not simply equal to the characteristic length 
$a_{\bot}$ of the transverse confinement. 

We observe that a generalized Lieb-Liniger 
action functional, which describes also the Tonks-Girardeau 
regime \cite{lieb}, where $g |\phi|^2 \ll \gamma^2$,  
and reduces to the NPSE under the condition 
$g |\phi|^2 \gg \gamma^2$, was derived in Ref. \cite{sala-lieb}. 
Clearly, in the NPSE regime one can distinguish two sub-regimes: 
the 1D quasi-BEC regime for $\gamma^2 \ll g |\phi|^2 \ll 1$ 
where $\sigma=1$ and the 3D BEC regime for $g |\phi|^2 \gg 1$ 
where $\sigma = (g |\phi|^2)^{1/4}$ \cite{sala-npse,sala-lieb}. 

\section{Numerical Results}

We perform the numerical simulation of Eq. (\ref{gp}) with (\ref{pot1}) 
employing real-time propagation  with 
Crank-Nicholson discretization scheme \cite{bo}. 
Because of the oscillating nature of the optical 
potential (\ref{pot1}) great care is needed 
to obtain a precise localized state. The accuracy of the numerical 
simulation has been tested by varying the space and time steps as well as the 
total number of space steps. We choose the initial condition 
\beq 
\phi(z,0) = {1\over \pi^{1/4} \eta^{1/2}} \ e^{-z^2/(2\eta^2)} \; 
\eeq
with $\eta=1$ and imposing vanishing boundary conditions 
$\phi(\pm z_B,t) = 0$ with $z_B = 100$. 
We stop the dynamics when a ``stationary'' configuration is reached. 
For $g=0$ we have numerically verified that a different choice of $\eta$ 
in the narrow Gaussian initial wavefunction 
does not affect the long-time behavior of the evolving wavefunction, 
which is definitely not Gaussian but a multi-peak localized configuration. 

\begin{figure}
\begin{center}
\includegraphics[width=\linewidth]{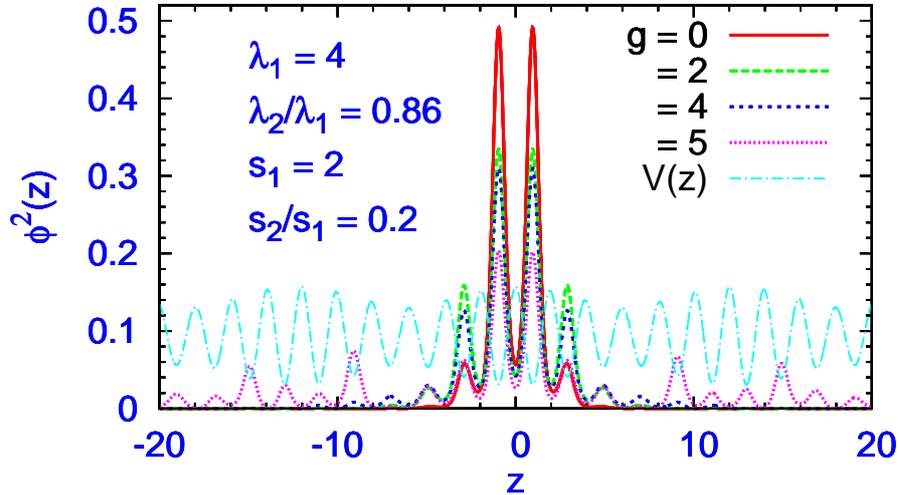}
\end{center}
\caption{Typical density distribution $\phi^2(z)$ vs. 
$z$ for an interacting BEC with different values of the 
interaction strength $g= 2Na_s/a_\perp$. 
The quasi-periodic optical-lattice potential $V(z)$, Eq. (\ref{pot1}) 
is plotted in arbitrary units with  $\lambda_1=4, 
\lambda_2/\lambda_1=0.86, s_1=2, s_2/s_1=0.2$.
Adapted from \cite{sala-loc}.}
\label{fig1}
\end{figure}

We study the effect of interaction in a BEC of $^{39}$K 
atoms with scattering length $a_s= 33a_0=1.75$ nm \cite{wang} (with 
$a_0=05292$ nm, the Bohr radius,) by solving Eq. (\ref{gp}) with 
potential (\ref{pot1}). In present dimensionless units 
this will correspond to $a_s/a_\perp= 0.00175$. The 
inclusion of the repulsive nonlinear potential term in Eq. (\ref{gp}) 
will reduce the possibility of the appearance of localized bound states. 

This is illustrated in Fig. \ref{fig1} where we plot the 
density distribution for $\lambda_1=4, \lambda_2/\lambda_1=0.86, s_1=2, 
s_2/s_1=0.2$ for potential (\ref{pot1})  
and different $g= 2N\hat a/a_\perp=(0,2,4,5)$. The figure shows 
the following remarkable results: 
i) for $g=0$ the localized state is confined between $z=\pm 10$; 
for $g=2$ the matter density is reduced in the 
central peaks and new peaks appear for larger $z$ values; iii) 
for $g=4$ the matter density is further reduced 
in the central region and new peaks appear in the form of ondulating tails
near the edges; iv) with further increase in the value of $g$, 
the localized states have larger and 
larger spatial extension and soon the nonlinear repulsion is so large 
that no localized states are possible and this happens rapidly as $g$ is 
increased beyond $5$. 

The nonlinearity in Eq. (\ref{gp}) is $g=2 a_s 
N/a_\perp$ and for about 1800 $^{39}$K atoms with $a_s=0.00175$ 
\cite{wang} the nonlinearity has the typical numerical value $g\approx 
6$. Such a small nonlinearity has a large effect on localization 
of a $^{39}$K BEC and prohibits the localization. However, the number 
of K atoms can be increased if the scattering length is 
reduced by varying an external background magnetic field near a Feshbach 
resonance \cite{fesh}.  

\begin{figure}
\begin{center}
\includegraphics[width=\linewidth]{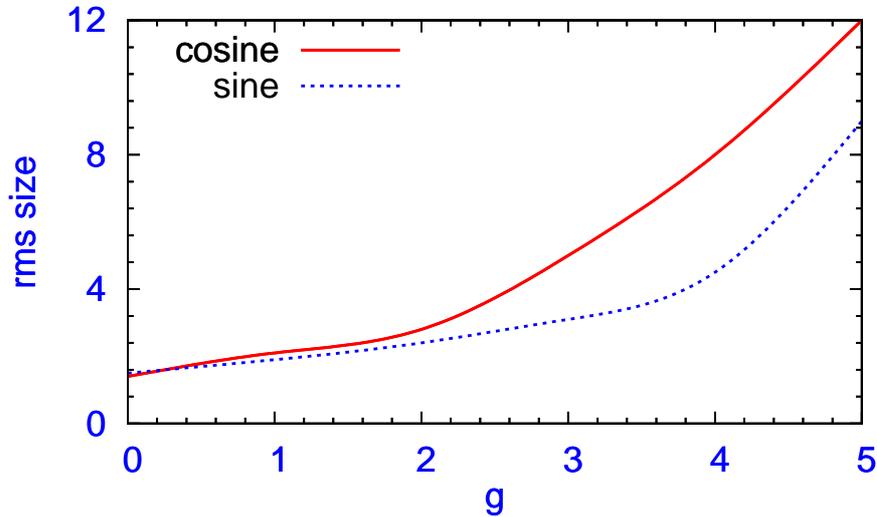}
\end{center}
\caption{The root-mean-square (rms) size vs. 
interaction strength $g$ of the BEC in the 
quasi-periodic potential of Eq. (\ref{pot1}) (cosine), and a similar 
potential where cosines are substituted by sines (sine),  
with $\lambda_1=4, \lambda_2/\lambda_1=0.86, s_1=2, s_2/s_1=0.2.$
All quantities plotted are in  dimensionless units. 
Adapted from \cite{sala-loc}.}
\label{fig2}
\end{figure}

As $g$ value is increased, the root mean square 
(rms) size of the BEC increases. For values of $g$ larger that $5$ 
the localization is fully suppressed, corresponding to 
the destruction of localization. 
The increase in the rms size of the localized state with the increase 
in $g$ is illustrated in Fig. \ref{fig2}, where we plot the rms size vs. $g$ 
for potential (\ref{pot1}), and also for a similar potential 
where cosines are substituted by sines. 

It should be noted that in the experiment of 
Roati {\it et al.} \cite{roati} the residual scattering length of 
$^{39}$K atoms near the Feshbach resonance was 0.1$a_0$ (= 0.0053 nm), 
e.g., they can vary the scattering length in such small steps. 
Thus it should be possible experimentally to obtain the 
curves illustrated in Fig. \ref{fig2} and compare them  with the 
present investigation. 

\section{Conclusion}

By numerically solving the 1D Gross-Pivaevskii 
equation (here derived from the many-body quantum field theory 
through a nonpolynomial Schr\"odinger equation)  
we have verified the phenomenon of localization for 
a non-interacting Bose-Einstein 
condensate in a quasi-periodic 1D optical-lattice potential 
prepared by two overlapping polarized standing-wave laser beams. 
However, we have found that a sufficiently large repulsive atomic 
interaction destroys the localization. In particular, 
we have investigated 
this effect by changing the strength $g= 2Na_s/a_\perp$ 
of the nonlinearity in the 1D Gross-Pivaevskii equation: 
as $g$ is gradually increased, the localization is slowly weakened 
with the localized state extending over a large space domain. 
Eventually, for $g$ greater than $5$ the localization is 
substantially suppressed.

\end{document}